\documentclass[letterpaper, 10 pt, conference]{ieeeconf}  

\IEEEoverridecommandlockouts                              

\usepackage{graphicx}
\usepackage{soul, color, xcolor}
\usepackage{amsmath,amssymb,amsfonts,}
\usepackage{caption}
\title{\LARGE \bf
How Collective Intelligence Emerges in a Crowd of People Through Learned Division of Labor: A Case Study}

\author{Dekun Wang and Hongwei Zhang*, \textit{Senior Member, IEEE}
\thanks{This work was supported by the Guangdong Basic and Applied Basic Research Foundation under project 2023A1515011981, and Shenzhen Science and Technology Program under project GXWD20231129102406001.}
\thanks{*LinYi's experiment can be found in: {\footnotesize bilibili.com/video/BV1Rd4y1R7tG and bilibili.com/video/BV1DB4y1N7QU}}
\thanks{The authors are with the Guangdong Provincial Key Laboratory of Intelligent Morphing Mechanisms and Adaptive Robotics, School of Mechanical Engineering and Automation, Harbin Institute of Technology, Shenzhen, Guangdong, 518055, P.R. China.}%
\thanks{* All correspondence should be addressed to H. Zhang. {\tt\small hwzhang@hit.edu.cn}}%
}

\begin{document}

\maketitle
\thispagestyle{empty}
\pagestyle{empty}

\begin{abstract}

This paper investigates the factors fostering collective intelligence (CI) through a case study of *LinYi's Experiment, where over 2000 human players collectively controll an avatar car. By conducting theoretical analysis and replicating observed behaviors through numerical simulations, we demonstrate how self-organized division of labor (DOL) among individuals fosters the emergence of CI and identify two essential conditions fostering CI by formulating this problem into a stability problem of a Markov Jump Linear System (MJLS). These conditions, independent of external stimulus, emphasize the importance of both elite and common players in fostering CI. Additionally, we propose an index for emergence of CI and a distributed method for estimating joint actions, enabling individuals to learn their optimal social roles without global action information of the whole crowd.

\end{abstract}

\section{INTRODUCTION}
\label{sec1}

Collective intelligence, also known as the wisdom of crowds, describes the phenomenon wherein groups of human individuals exhibit intelligent behavior in opinion formation, decision-making and multiplayer games among others \cite{od3, epm2}. Multiplayer games, such as Sharing Control Games (SCG), offer scientists a controlled testbed to investigate the essential factors fostering collective intelligence \cite{scg1}. Understanding these factors is crucial for enhancing group performance and addressing societal challenges.

\begin{figure}[thpb]
    \centering
    \includegraphics[width=0.49\textwidth]{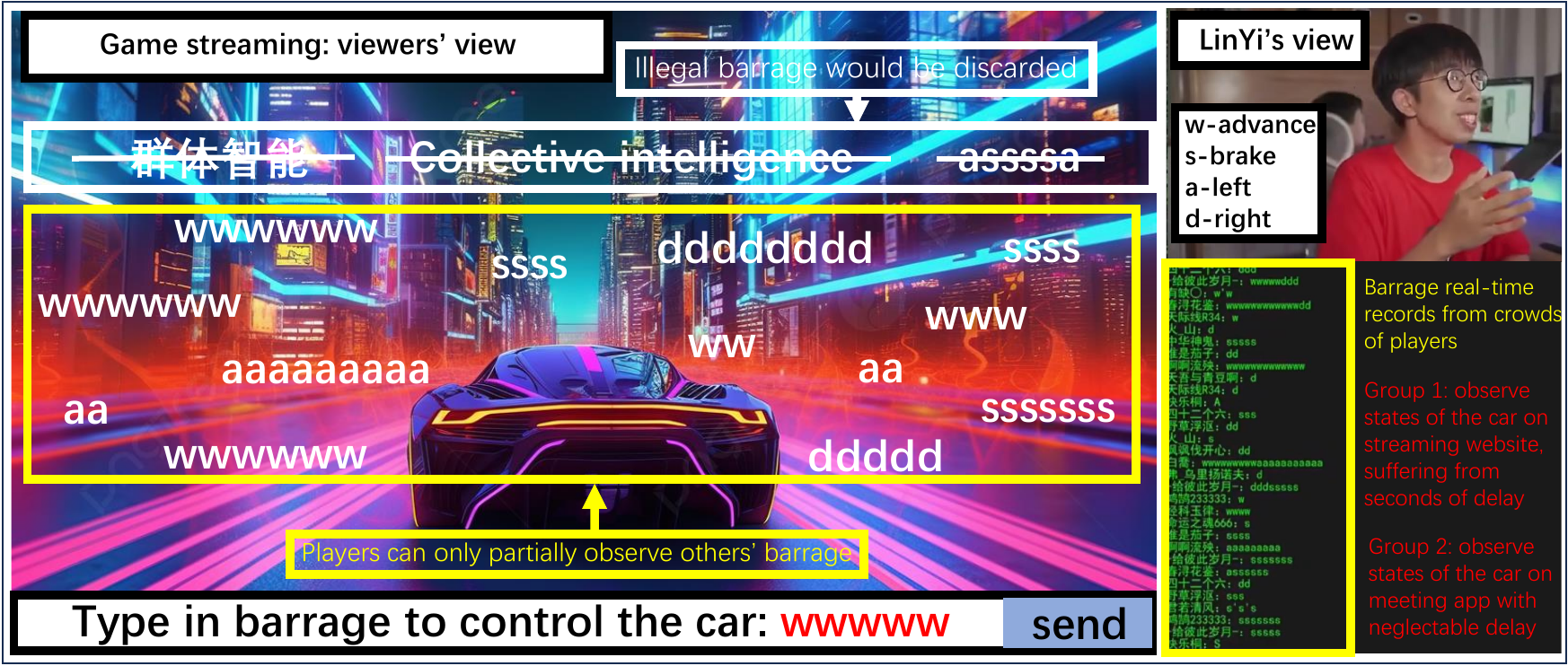}
    \caption{LinYi's Sharing Control Game: $N$ players share control over one single car. `w',`s',`a',`d' stand for `advance',`brake',`left',`right' respectively. In this paper, `advance' or `brake' are represented by throttle $T$,`left' or `right' are done by steering angle $\delta$. Differences between 'wwww' and 'w' are captured by different magnitudes of throttle $T$.}
    \label{fig1}
\end{figure}

In 2022, a popular SCG experiment was conducted by a streamer named *LinYi on the bilibili streaming platform, where over 2000 human players collectively control the motion of an avatar car (Fig.\ref{fig1}). Divided into two groups based on capabilities, players initially assume random roles and continually adjust role policies by experiences. Over time, a spontaneous division of labor emerges, with one group solely control the throttle and the other steering, thereby effectively maneuvering the car. This experiment exemplifies the emergence of CI within a specific crowd of people, yet the factors fostering CI remain unclear. Our paper aims to pinpoint these factors through a case study of this experiment.

To date, the factors fostering CI, studied across disciplines including opinion dynamics, game theory and societal experiments, can be categorized into three main domains: 1) communication, 2) regulation and constraints, 3) cooperation. 

Communication serves as a foundation for other two domains. Studies in opinion dynamics have mathematically revealed that greater social power \cite{od1}, susceptibility, network centrality \cite{od3}, and appropriate levels of stubbornness \cite{od4} for elite individuals contribute to CI. Furthermore, experimental studies involving crowds of people have also demonstrated the significance of social power, susceptibility \cite{epm4} and performance feedback mechanism \cite{epm2, epm3}. These factors are closely tied to the properties of the communication graph. 

Regulation and constraints are typically enforced by a central authority, which incentivizes desired behaviors and discourage unwelcome ones. In the field of game theory \cite{gd1}, insights into collective intelligence reveal challenges when individual interests diverge from group goals. Rational individuals, driven by self-interest, may pursue strategies that ultimately undermine group interests, i.e., a certain index of CI. Regulations like laws and social norms are introduced to mitigate this dilemma. Studies in opinion dynamics also underscore the significance of such aspects, namely social pressure \cite{od6} and logical constraints \cite{od5}. 

The significance of cooperation in fostering CI has long been recognized as the saying goes, ``two heads are better than one." Specifically, cooperation requires collaboration, coordination, reciprocity \cite{gd1}. Notably, division of labor, a special aspect of cooperation, describes individuals selecting roles based on their capability disparities and comparative advantages. This process is often spontaneous and decentralized through experiential learning. However, all existing studies on CI lack learning mechanisms for modelling such self-organized learning process. 

Multi-agent reinforcement learning (MARL) offers a compelling framework for modelling the learning process of social roles and division of labor within human groups \cite{rl}. Role-based method \cite{marl2} represent agents' specified behaviors as social roles, demonstrating how agents learn appropriate social roles based on different capabilities. Zhang et al. \cite{gd3} propose fully decentralized approaches for cooperative MARL, empowering each individual to optimize its local policy towards optimizing global returns in a decentralized manner, i.e., using solely local return information. 

However, MARL frameworks implicitly assume that division of labor stems solely from individuals' pursuit of maximizing external rewards \cite{rl}, overlooking what else inherent imperative within a crowd that drives this division, as suggested by the concept of intrinsic motivation in psychology \cite{psychology}. This oversight is also seen in opinion dynamics, game theory, and societal experiments, necessitating further investigation. Furthermore, we observed that individuals lack full access to the role information of others during the learning process in LinYi's experiment, requiring a fully distributed MARL method to model the process.

This paper aims to identify essential factors fostering CI through a case study of LinYi's experiment. We propose a general SCG model and validate this model by replicating observed behaviors in the experiment. This model reveals how individuals spontaneously establish division of labor through experiential learning, ultimately leading to CI. Moreover, we identify other conditions fostering CI, which reveal that CI only emerges when the following two thresholds are met: 1) the total number of individuals, and 2) the proportion and social power of elite individuals. Moreover, we show the imperative of these conditions stems from the stability property of a system. Additionally, we propose an index for emergence of CI and a method for estimating joint actions, enabling individuals to learn optimal roles without global action information. Furthermore, these findings are validated through numerical simulations.

\textit{Notations:} We denote spaces as calligraphy $\mathcal{X}$, matrix as $X$, vector and scalar as $x$. Let $[N]=\{1, 2,...,N\}$, $\mathcal{Z} = \{0,1,2,...\}, \mathcal{Z}^{+} = \mathcal{Z}\setminus\{0\}$, the cardinality of a finite set $\mathcal{X}$ be $|\mathcal{X}|$, and square matrix with same rows be $\mathcal{SM}[row]$. Note that we sometimes omit the independent variables of a function for brevity, e.g. $\kappa^i(x(k))$ as $\kappa^i$, without causing any confusion.

\section{CROWD DECISION-MAKING MODEL}
\label{sec2}

\subsection{General Modelling: the SCG model}
To model SCG, we first need to distinguish it from typical control systems. Consider the following discrete-time system with a sampling period $t_s$:
\begin{equation} 
\label{eq1}
    x(k+1) = f(x(k),u(k)),\ x\in\mathbb{R}^n, u\in\mathbb{R}^m
\end{equation}

In typical control problems, it is often assumed that the exact model of system (\ref{eq1}) is known. This knowledge allows for the design of control policies aimed at guiding the system to behave as desired. Specifically, let us consider a control objective $\mathcal{O}$: tracking a given reference under constraints. An effective control policy $\kappa^+$ can be designed to achieve $\mathcal{O}$ using tools from control theories. By indexing the time instance $0, t_s, 2t_s,...$ as step $k\in\mathcal{Z}$, the closed-loop system behaves as follows: at time step $k$, control commands $u(k)$ are input into the system (\ref{eq1}), where $u(k)$ is determined by the policy $\kappa^+$ for any $k\in\mathcal{Z}$.

In a SCG where N players collectively control the system (\ref{eq1}), any players have a learning mechanism of social roles and the capability, i.e., control policy $\kappa^+$ prior to learning of roles to accomplish $\mathcal{O}$. However, each player can only input its control commands to one element of $u(k)$, denoted as $u_i(k)\in\mathbb{R}$ for some periods $T_s>>t_s$, as observed in LinYi's experiment. We refer to one player controlling element $u_i(k)$ as it takes social role $i (i\in[m]$). Over time, each player tries different roles many times and eventually converges to a certain role.

\textit{Remark 1:} It is thus evident that system (\ref{eq1}) with control policy $\kappa^+$ under typical control problem setting represents the optimal outcome that the corresponding SCG can hope to achieve. Therefore, this optimal scenario is denoted as the "Baseline case under $\kappa^j$", $j\in\{-,+\}$,  serving as the baseline for SCG cases.

Before proceeding, we make the following assumptions:

\textit{Assumption 2 (Accessible Information): } a). Players have complete access to plant states regardless of delays. Throughout this paper, we focus on a scenario with two groups: group 1 with $N_1$ players, experiences the same delay of $\tau t_s$, while group 2 with $N_2$ players has no delay. b). Players are aware only of the role selections of their neighbors in a communication graph $\mathcal{G}_k$, as Fig.\ref{fig2}(b) shows.

Without loss of generality, players from each group can be indexed as $\mathcal{N}_1=[N_1]$ and $\mathcal{N}_2 = [N]\setminus\mathcal{N}_1$ respectively, where $N=N_1+N_2$. 

\textit{Assumption 3 (Homogeneity of Players): } a). All players share a common control policy $\kappa^+$. b). players within a group have identical social powers $\rho_s\in\mathcal{Z}^+$, which represents relative significance of decisions made by certain players \cite{od3}. Specifically, players in $\mathcal{N}_2$ have social power $\rho_s\geq1$, while those in $\mathcal{N}_1$ have social power of $1$, which will not be explicitly denoted.

It should be noted that due to information delays, commands taken by players in $\mathcal{N}_1$ with $\kappa^+$ in fact function as another policy, denoted as $\kappa^-$. How $\kappa^-$ is derived directly from $\kappa^+$ will be presented in subsection \ref{sec2}.C. and section \ref{sec3}. 

\textit{Assumption 4 (Control Periods of Players): } As mentioned above, each player can only input its control commands to $u_i(k)$ in a SCG for some periods $T_s$. We assume these $T_s$ are identical for any players, where $T_s=Kt_s, K\in\mathcal{Z}^+$. Thus $Kt_s$ is also the period of players taking their roles. Specifically, players with $\rho_s\neq1$ can control the same element $u_i(k)$ at most $\rho_s$ times within $K$ time steps.

Therefore, under Assumptions 2, 3 and 4, SCG can be formulated as a variant of Networked Multi-Agent MDP as defined in \cite{gd3}, which is characterized as a tuple $(\mathcal{X},\{\mathcal{A}^i\}_{i\in [N]},P,R,\{\mathcal{G}_{t_j}\})$ where $\mathcal{X}$ is the state space of system (\ref{eq1}) shared by all the players, $\mathcal{A}^i=[m]$ is the action space of player $i$ (action means to select a role), $\mathcal{G}_{t_j}$ is a time-varying communication network, where $t_j = k+jK, j\in\mathcal{Z}$. Let $\mathcal{A}=\prod^{N}_{i=1}\mathcal{A}^i$ the joint action space. Then, $R : \mathcal{X}\times\mathcal{A}\rightarrow\mathbb{R}$ is a common reward function for all players, which quantifies operating performance of system (\ref{eq1}). In addition, $P:\mathcal{X}\times\mathcal{A}\times\mathcal{X}\rightarrow[0,1]$ is the state transition probability of the MDP. It should be noted that the control policies $\kappa^+$ and $\kappa^-$ are included in $P$, thus the learning of roles is not a completely model-free process, as illustrated in Fig.\ref{fig2}.(a).

At time step $t_j$, each player selects its own role $a^i(t_j)$, given state $x(t_j)$ or $x(t_j-\tau)$, following a local role policy $\pi^i:\mathcal{X}\times\mathcal{A}^i\rightarrow[0,1]$, which represents the probability of selection role $a^i(t_j)$ given state information $x$. Notably, all of local role policies form a joint policy $\pi: \mathcal{X}\times\mathcal{A}\rightarrow[0,1]$, which satisfies $\pi(x,a)=\prod^{}_{i\in[N]}\pi^i(x,a^i)$. All of role selections form a joint role selection $a(t_j)=[a^1(t_j),...,a^N(t_j)]^T$. Since the joint action, i.e., global role information of all players is not accessible to any players, player $i$ instead estimates $a(t_j)$ by $\underline{\hat{a}}^i(t_j)$ for any $i\in[N]$, which will presented in subsection \ref{sec2}.B. With accessible information player $i$, i.e., $<x(t_j), \underline{\hat{a}}^i(t_j), r(t_{j+1}), x(t_{j+1})>$ or $<x(t_j-\tau), \underline{\hat{a}}^i(t_j), r(t_{j+1}), x(t_{j+1}-\tau)>$, it updates its role policy $\pi^i$ to maximize $J_1$, i.e., Eq.(\ref{eq2}) in subsection \ref{sec2}.B. Therefore, we note this model is fully distributed.

\subsection{Learning Mechanism of Social Role}
As we point out that SCG can be treated as a Networked Multi-Agent MDP, it is thus evident that learning process of social roles can be modelled by MARL method. Combining concensus-based(Algorithm 1. \cite{gd3}) and role-based method, with a joint role selection estimation, we propose the following learning mechanism under actor-critic framework:

For any agent $i$, we assume that the role policy $\pi_{\theta^i}^i$ is parameterized by $\theta^i\in\mathbb{R}^z$. Then we have joint role policy $\pi_{\theta}(x(t_j),a(t_j))$ with joint parameter $\theta=[(\theta^1)^T,...,(\theta^N)^T]^T\in\mathbb{R}^{Nz}$. 

The objective of all agents is defined as follows:
\begin{equation} 
\label{eq2}
    \mathop{max}\limits_\theta J_1(\theta) = \sum\limits_{t_j=0}\gamma^{(j-1)}\mathbb{E}(r(t_j))
\end{equation}
where $\gamma$ is the discount factor. 

\textit{Definition 5 (CI under SCG):} Collective intelligence emerges when $J_1(\theta)\geq J_0$, where we denote the expected return of ``Baseline case under $\kappa^+$" as $J_0$. 

\begin{figure}[thpb]
    \centering
    \includegraphics[width=0.49\textwidth]{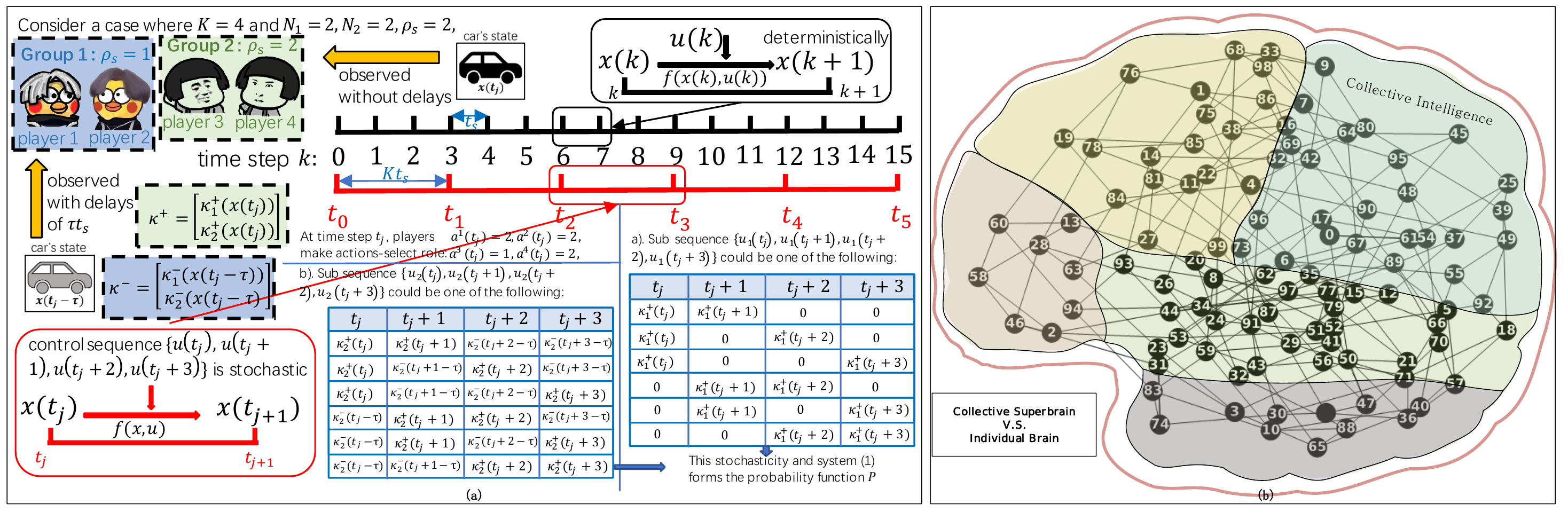}
    \caption{(a): Flow chart of a SCG example. Division of labor within human societies reflect a natural SCG: in a hospital, doctors are assigned to different departments according to their specialties, a 'spatial distribution'. Doctors within a department rotate in shifts, a 'temporal distribution'. (b): One example of the random communication graph $\mathcal{G}_k$, which is connected, undircted, sparse graph with $N$ nodes.}
    \label{fig2}
\end{figure}

The global action-value function under policy $\pi_{\theta}$ becomes
\begin{equation} 
\label{eq3}
    Q_\theta(x,a) = \sum\limits_{t_j}\mathbb{E}[r(t_j)|x_0=x,a_0=a,\pi_{\theta}]
\end{equation}
and the global state-value function $V_{\theta}(x)$ is defined as $V_{\theta}(x) = \sum_{a\in\mathcal{A}}\pi_{\theta}(x,a)Q_{\theta}(x,a)$. 
Moreover, we define a local advantage function as follows:
\begin{equation} 
\label{eq4}
    A^i_{\theta}(x,a)=Q_{\theta}(x,a)-\tilde{V}^i_{\theta}(x,a^{-i}),
\end{equation}
\begin{equation} 
\label{eq5}
    \tilde{V}^i_{\theta}(x,a^{-i})=\sum_{a^i\in\mathcal{A}^i}\pi^i_{\theta^i}(x,a^i)\cdot Q_{\theta}(x,a^i,a^{-i})
\end{equation}
where $a^{-i}$ is the joint role selection except for agent $i$. We assume each player has its own estimation $Q_\theta(x,a,\omega^i)$ parameterized by $\omega^i\in\mathbb{R}^h$, by Eq.(\ref{eq4}), (\ref{eq5}) estimation of $A^i_\theta(x,a)$ is obtained. By Theorem 3.1. from \cite{gd3}, the gradient of $J_1(\theta)$ with respect to $\theta^i$ is given by
\begin{equation} 
\label{eq6}
    \nabla_{\theta^i}J_1(\theta) = \mathbb{E}_{x\sim d_{\theta},a\sim \pi_{\theta}}[\nabla_{\theta^i}log\pi ^i_{\theta^i}(x,a^i)\cdot A_{\theta}^i(x,a)]
\end{equation}
Each agent $i$ shares the local parameter $\omega^i$ with its neighbors on the network $\mathcal{G}_{t_j}$ (Fig.\ref{fig2} (b)) with which information is aggregated with a weight matrix $C_{t_j}=[c_{t_j}(i,j)]_{N\times N}$. The weight matrix is given by Eq.(3) from \cite{naomi}.

\textit{Remark 6:} Given \textit{Assumption 2} and $\mathcal{A}^i=[m]$, we observe that the joint role selection $a(t_j)\in\mathbb{R}^N$ can be completely represented as $\underline{a}(t_j)\in\mathbb{R}^2$, (i.e. $\mathbb{R}^N$ projected to $\mathbb{R}^2$), if $m=2, N\geq2$, and $N_1, N_2$ is known to players. Specifically, if $n_{ij}$ denotes as the number of players selecting role $j$ from group $i$, $\underline{a}(t_j)=[n_{11},n_{21}]$ can fully capture the information of $a(t_j)$. Therefore, unlike algorithms in \cite{gd3} that necessitate global information (the role selections of all players), we can not only estimate $\underline{a}(t_j)$ using solely local information, but also significantly reduces complexities of actor-critic networks. 

The estimation method is designed as follows: at time $t_j$, player $i$ has a neighbour set $\Lambda^i\subseteq[N]$ determined by communication graph $\mathcal{G}_{t_j}$. All the neighbours $j\in\mathcal{N}_1$ construct a subset $\Lambda^i_1$ while those $j\in\mathcal{N}_2$ construct another subset $\Lambda^i_2$. Given \textit{Assumption 2}, neighbours who take role action 1 construct another subset $\Lambda^i_{r,1}$ within $\Lambda^i_{r}, r\in{1,2}$. Denote $a^i(t_j)$ as $a^i_{t_j}$, the agent $i$'s estimation is:
a) For player $i\in\mathcal{N}_1$:
\begin{equation} \label{eq7}
\hat{\underline{a}}^{i}(t_j)=\left\{
        \begin{array}{ll}
            [N_1\frac{|\Lambda^i_{1,1}|+1}{|\Lambda^i_1|+1},N_2\frac{|\Lambda^i_{2,1}|}{|\Lambda^i_2|}]^T &\text{if } a^i_{t_j}=1  \\
            {[N_1\frac{|\Lambda^i_{1,1}|}{|\Lambda^i_1|+1},N_2\frac{|\Lambda^i_{2,1}|}{|\Lambda^i_2|}]^T}  &\text{if } a^i_{t_j}=2 \\
        \end{array}
    \right.
\end{equation}
b) For player $i\in\mathcal{N}_2$, $\hat{\underline{a}}^{i}(t_j)$ is estimated similarly.
c) For any case that denominators in a), b) are zero, assume the corresponding element of $\hat{\underline{a}}^{i}(t_j)$ as $0.5N_r$, $r\in\{1,2\}$, i.e. half the group take role $1$.

Therefore, we introduce learning mechanism of social role. The critic network iterates as:
\begin{gather}
    \label{eq8}\delta_{t_j}^i = r(t_{j+1})+Q_{t_{j+1}}(\omega_{t_j}^i)-Q_{t_j}(\omega_{t_j}^i)\\
    \label{eq9}\tilde{\omega}_{t_j}^i=\omega_{t_j}^i+\beta_{\omega,{t_j}}\cdot \delta_{t_j}^i\cdot\nabla_\omega Q_{t_j}(\omega^i_{t_j})\\
    \label{eq10}\omega_{t_{j+1}}^i = \sum_{j\in \mathcal{N}} c_{t_j}(i,j)\cdot\tilde{\omega}_{t_j}^j
\end{gather}
The actor network iterates as:
\begin{gather}
    \label{eq11}A_{t_j}^i=Q_{t_j}(\omega_{t_j}^i)-\sum_{a^i\in \mathcal{A}^i}\pi_{\theta^i_{t_j}}^i(x_{t_j},a^i)\cdot Q_{t_j}(\omega_{t_j}^i)\\
    \label{eq12}\psi_{t_j}^i=\nabla_{\theta^i}log\pi^i_{\theta^i_{t_j}}(x_{t_j},a_{t_j}^i)\\
    \label{eq13}\theta_{t_{j+1}}^i = \theta_{t_j}^i + \beta_{\theta,{t_j}}\cdot A_{t_j}^i\cdot\psi_{t_j}^i
\end{gather}
It is worth-noting that $Q_{t_{j+1}}(\omega_k^i)$ stands for $Q_{t_{j+1}}(x(t_{j+1}),\underline{\hat{a}}(t_{j+1}),\omega_{t_j}^i)$, unlike that in \cite{gd3}. Moreover, $\sum_{i=1}^{N}\hat{\underline{a}}^{i}(t_j)/N\rightarrow\underline{a}(t_j)$ as $N\rightarrow\infty$, thus the estimation method fits consensus-based algorithms Eq.(\ref{eq8})-(\ref{eq13}).

\subsection{Modelling of Brought-in Capabilities}
Capabilities prior to learning of social roles are represented as control policies, designed using typical model-based control methods. 

For example, to design an MPC controller, we usually solves following optimization problem at time $k$:
\begin{align} 
\label{eq14}
\vspace{-0.5cm}
J_2 = &  \sum_{k=1}^{N_p} [x^T(k)-x_{ref}^T(k)] Q [x(k)-x_{ref}(k)]  \nonumber \\
& +[u^T(k)-u_{ref}^T] R [u(k)-u_{ref}]
\end{align}
where $Q$, $R$ are the positive semi-definite weight matrices, $x_{ref}, u_{ref}$ are predefined references, the cost objective Eq.(\ref{eq14}) subjects to a linear time-varying system:
\vspace{-0.2cm}
\begin{equation}
\vspace{-0.3cm}
\label{eq15}
    x(k+1) = A_kx(k) + B_ku(k) + d_k
\end{equation}
and the following constraints:
\vspace{-0.2cm}
\begin{equation}
\vspace{-0.2cm}
\label{eq16}
    u(k)\in\mathcal{U}_1, \Delta u(k)\in\mathcal{U}_2, x(k) \in \mathcal{X}
\end{equation}

At time $k$, the MPC controller uses the first element of the solution to problem (\ref{eq14}) as the input to the system, forming a control policy denoted as $\kappa(x(k))$. However, if state information is delayed by $\tau t_s$, it results in another policy, denoted as $\kappa(x(k-\tau))$. For simplicity and coherence, we denote $\kappa(x(k)$, $\kappa(x(k-\tau))$ as $\kappa^{+}$ and $\kappa^{-}$ respectively, as presented in subsection \ref{sec2}.A. 

\section{Theoretical Analysis}
\label{sec3}
In this section, we first introduce the modeling of the avatar car, then a simplified model is presented for the convenience of theoretical analysis. 

Denote the avatar car's global positions as $p_x, p_y$, yaw angle $\theta$, longitudinal velocity $v$, throttle $T$, steering angle $\delta$, the length of car $L$, the acceleration factor $\alpha$, and a constant drag force $F$, the dynamics can be modeled as:
\vspace{-0.2cm}
\begin{align}
    \label{eq17}&\dot{p_x} = cos(\theta)\cdot v \\
    \label{eq18}&\dot{p_y} = sin(\theta)\cdot v \\
    \label{eq19}&\dot{\theta} = tan(\delta)/L\cdot v \\
    \label{eq20}&\dot{v} = -F + \alpha\cdot T
\end{align}

Before replicating LinYi's Experiment using simulation, we examine a simplified case in this section. We only take Eq.(\ref{eq19})-(\ref{eq20}) and linearize them around nominal points. When nominal points $\theta_0, \delta_0 \approx 0$, we obtain a simplified model:
\begin{align}
    \label{eq21}&\dot{\theta} = V_0/L\cdot\delta \\
    \label{eq22}&\dot{v} = \alpha\cdot T
\end{align}

The objective is to guide system(\ref{eq21})-(\ref{eq22}) to track reference trajectory $\theta_{ref}$ and $v_{ref}$. We consider $\dot{\theta}_{ref}=\dot{v}_{ref} = a_{cc}$. We can get an error dynamics: $\dot{e}_{\theta} = \dot{\delta} - \dot{\delta}_{ref} = \frac{V_0}{L}\cdot\delta - a_{cc}$, $\dot{e}_{v} = \dot{v} - \dot{v}_{ref} = \alpha\cdot T - a_{cc}$.

With zero input, we observe both $e_\theta$ and $e_v$ diverge linearly. Using numerical methods, we acquire an exponential factor $\lambda$ that best approximates the linear divergence within 10 seconds, resulting in an approximated system:
\begin{align}
    \label{eq23}&\dot{e}_{v} \approx \lambda\cdot e_{v} + \alpha\cdot T\\
    \label{eq24}&\dot{e}_{\theta} \approx \lambda\cdot e_{\theta} + V_0/L\cdot\delta
\end{align}

With state $x=[e_v,e_\theta]^T$ and input $u=[T,\delta]^T$, we give a reward function $R=-||x(t)||$ and rewrite system Eq.(\ref{eq23})-(\ref{eq24}) as:
\vspace{-0.3cm}
\begin{equation} 
\vspace{-0.3cm}
\label{eq25}
    \dot{x}(t)=g(x(t),u(t))
\end{equation}

Since system (\ref{eq25}) is decoupled, we denote subsystems as $\dot{x}_i(t)=g_i(x_i(t),u_i(t))$, where $x_i(t), u_i(t)$ refer to $i$-th element of $x(t), u(t)$, for $i\in\{1,2\}$ corresponding to Eq.(\ref{eq23}),(\ref{eq24}) respectively. 

Therefore, the reference tracking of system Eq.(\ref{eq21})-(\ref{eq22}) is approximated as stabilization of system (\ref{eq25}). Then we can design a state feedback policy $\kappa^+=-K_1x(t)$ to stabilize system (\ref{eq25}), where $K_1=[k_{11},k_{12}]^T\in\mathbb{R}^2$. Assuming $x(t-\tau)\approx x(t)-\tau \dot{x}(t)$, another policy is derived as $\kappa^-=-K_1x(t-\tau t_s)\approx -K_2x(t)$, where $K_2=[k_{21},k_{22}]^T$. Additionally, the zero-input case is denoted as $\kappa^0=[0,0]^T$. We denote `sub-policy', the $i$-th element of policy $\kappa^j$ as $\kappa_i^j$ for $i\in\{1,2\},j\in\{-,0,+\}$. Then the subsystem $\dot{x}_i(t) = g_i(x_i(t),u_i(t))$ is controlled by switching policy $\{\kappa_i^-,\kappa_i^0,\kappa_i^+\}$. We denote the closed-loop subsystem under policy $\kappa_i^j$ as $\Gamma_i^j, j\in\{-,0,+\}$ for brevity. Therefore, after discretization, both closed-loop subsystems become a discrete-time Markov Jump Linear System(MJLS):
\vspace{-0.2cm}

\begin{equation} 
\label{eq26}
\mathcal{S}_i =  \left\{
        \begin{array}{ll}
            x_i(k+1) = \Gamma_{\psi_i(k)}x_i(k)  \\
            x_i(0) = x_{i0}, \psi_i(0)=\psi_{i0} \\
        \end{array}
    \right.
\end{equation}
where $\psi_i(k)$ represents a Markov chain taking values in $[N_a]$ which stands as an index for which subsystem we switch into, initial states and initial index $x_{i0}, \theta_{i0}$, $\Gamma_i=(\Gamma_i^-, \Gamma_i^0, \Gamma_i^+)\in\mathbb{H}^n$, where $\mathbb{H}^n$ is the linear space made up of all $N_a$ sequence of matrices $V=(V_1,V_2,...,V_{N_a})$ with $V_i\in\mathbb{R}^n$. Specifically, here $n=1$ and $N_a=3$.

Most importantly, as $p_{lv}$ represents the probability of transition from subsystem $l$ to subsystem $v$, the transition probability matrix $P_i=[p_{lv}]$ of subsystem $i$ is decided by role selections of players. Thus it varies for every $Kt_s$. However, when the division of labor is established, $\pi^i(x,1)$ converges to a consensus value $q$ for $i\in\mathcal{N}_1$, or to value $m$ for $i\in\mathcal{N}_2$ respectively(i.e. convergence of Algorithm 1, which is proved in\cite{gd3}), $P_i$ becomes constant, that is: 
\begin{equation} 
\label{eq27}
    P_i \triangleq\mathcal{SM}[\frac{K-\mathbb{E}(n_{1i}) - \mathbb{E}(n_{2i})}{K}, \frac{\mathbb{E}(n_{1i})}{K},\frac{\mathbb{E}(n_{2i})}{K}]
\end{equation}
where $\mathbb{E}(n_{11}) = qN_1, \mathbb{E}(n_{12}) = (1-q)N_1, \mathbb{E}(n_{21}) = m\rho_sN_2, \mathbb{E}(n_{22}) = (1-m)\rho_sN_2$, $n_{ij}$ for $i,j\in\{1,2\}$ is defined in \textit{Remark 6}.

Therefore, by following lemma, we can analyze $P_i$ under what conditions ensures both states of system (\ref{eq25}) converges to 0 with probability 1 (\textit{w.p.1}).

\textit{Lemma 7:} By Theorem 3.9 and Corollary 3.46. in \cite{mjls}, subsystem $\mathcal{S}_i$ (\ref{eq26}) is mean square stable (MSS), if and only if its spectral radius of $\mathcal{A}_{i1}$, $\sigma(\mathcal{A}_{i1})<1$. Moreover, if $\mathcal{S}_i$ (\ref{eq26}) is MSS, then $x_i(k)\rightarrow0$ \textit{w.p.1} as $k\rightarrow\infty$.

\begin{gather}
    \label{eq28}\mathcal{C}_i \triangleq P_i^T\otimes I_{n^2}\in\mathbb{R}^{N_an^2}\\
    \label{eq29}\mathcal{N}_i \triangleq diag[{\Gamma_i^j}^T\otimes\Gamma_i^j]\in\mathbb{R}^{N_an^2}\\
    \label{eq30}\mathcal{A}_{i1} \triangleq\mathcal{C}_i\mathcal{N}_i
\end{gather}

With $a_{cc}=30$, $\lambda=0.084$, and other parameters defined in Table \ref{table1}, we obtain $\Gamma_1^-,\Gamma_1^0,\Gamma_1^+= 0.996, 1.0017, 0.9997$ and $\Gamma_2^-,\Gamma_2^0,\Gamma_2^+= 1.053, 1.0017, 0.9897$. Therefore, it is evident that division of labor transfers subsystem $\mathcal{S}_i$ into deterministic asymptotic stable subsystems i.e. $\mathcal{S}_1\Rightarrow \dot{x}_1(t)=\Gamma_1^-x_1(t)$, $\mathcal{S}_2\Rightarrow \dot{x}_2(t)=\Gamma_2^+x_2(t)$.

\textit{Remark 8:} By analyzing the transition probability matrix, we establish the system (\ref{eq25}) converge to 0 with probability 1 under two conditions: (a) $\frac{N_1 + \rho_s N_2}{K}\geq0.95$, and (b) $\frac{\rho_sN_2}{N_1}\geq11.8\%$. The first condition explains why collective intelligence emerges only when the number of players reaches a certain threshold, highlighting the importance of allocating more social power to elite players, as observed in \cite{od3,epm4}. The second condition underscores the significance of the population proportions of elite individuals. These conditions are inherent factors for fostering collective intelligence and are independent of external rewards. Moreover, we establish a connection between the stability property of a MJLS and the imperative of DOL for fostering CI within players. This illustrates that the emergence of DOL and CI is driven not only by external stimulus but also by an inherent property.

\section{Numerical Simulation}
\label{sec4}
The objectives of players in LinYi's Experiment (Fig.\ref{fig1}) are in fact an obstacle avoidance reference tracking. To illustrate our work more clearly, we replicate observed behaviors in the experiment using numerical simulations. With $x=[p_x,p_y,\theta,v]^T, u=[T,\delta]^T$ and a given sample time $t_s$, Eqs.(\ref{eq17})-(\ref{eq20}) can be linearized into system (\ref{eq15}) after discretization. We specifically define constraints (\ref{eq16}) as $|u(k)|\leq u_{max}, |\Delta u(k)|\leq \Delta u_{max}$ and $H_{e} x(k) \leq G_{e}^k$(`Environment Envelop' as defined in \cite{wdk}, which consider obstacles, road boundaries as constraints). With $x_{ref}=[0,0,0,0]^T, u_{ref}=[F/\alpha,0]^T$ and cost function (\ref{eq14}), an obstacle avoidance MPC controller is derived, denoted as policy $\kappa^+$, the policy with information delay denoted as $\kappa^-$.

\subsection{Simulation Configurations}

\begin{table}[thpb]
    \centering
    \caption{Parameter Configuration}
    \begin{tabular}{|c|c||c|c|}
        \hline
        Parameters & Value & Parameters & Value\\ [0.5ex]
        \hline
         $a\_lr,c\_lr,\gamma$&  $10^{-4}, 10^{-2}, 0.9$ &  $K,\tau$& 80,100\\
         $N_{p},N_{c},t_s$&  2, 60, 0.02&  $Q$& diag[0,1,0,1]\\
         $u_{max}$&  $[\infty,\infty]^T$&  $R$& diag[0.1,0.1]\\
         $\Delta u_{max}$&  $[30,\frac{\pi}{30}t_s]^T$&  $L,W$& $5, 2$\\
         $d_{max},d_{min}$ & $10, 1$& $l_f,l_r,\alpha$& $2.5, 2.5, 0.5$\\
         \hline
    \end{tabular}
    \label{table1}
\end{table}
\begin{figure}[thpb]
    \centering
    \includegraphics[width=0.45\textwidth]{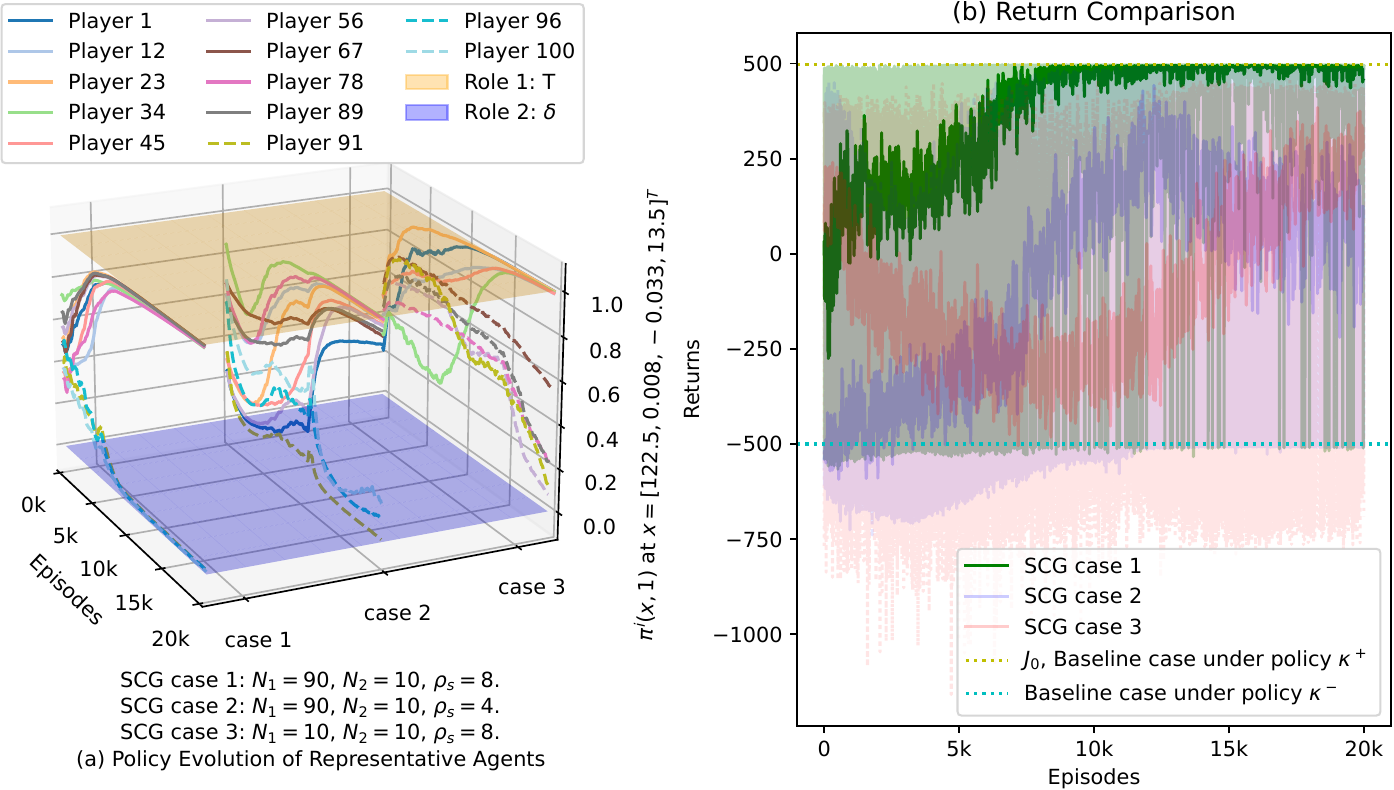}
    \caption{(a): role policy $\pi^i(x,1)$ evolution for three SCG cases. Specifically, players from group 2 are marked with dashed lines. (b): return vs episodes comparison between three SCG cases and baseline cases.}
    \label{fig3}
\end{figure}
\begin{figure}[thpb]
    \centering
    \includegraphics[width=0.48\textwidth]{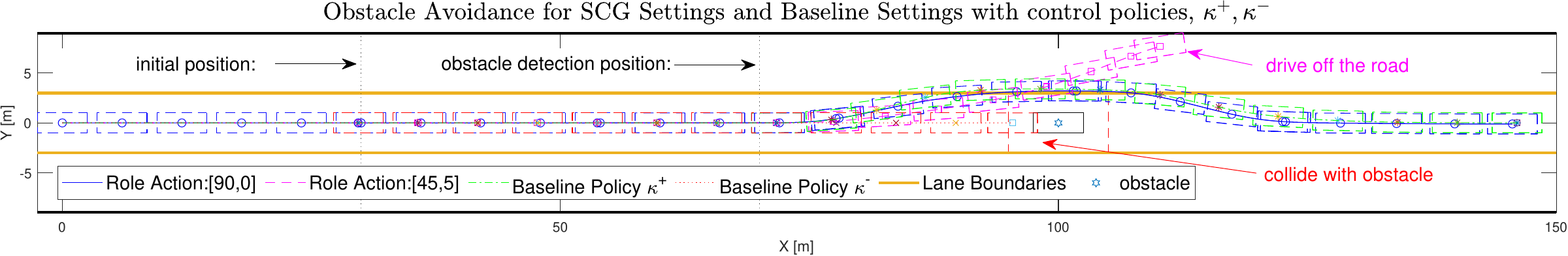}
    \caption{Trajectory comparison between SCG Case 1 and baseline cases. The joint role action $\underline{a} = [45,5]$ and $[90,0]$ can represent the group decisions when learning process just begins and that when learning process finished.}
    \label{fig4}
\end{figure}
\begin{figure}[thpb]
    \centering
    \includegraphics[width=0.48\textwidth]{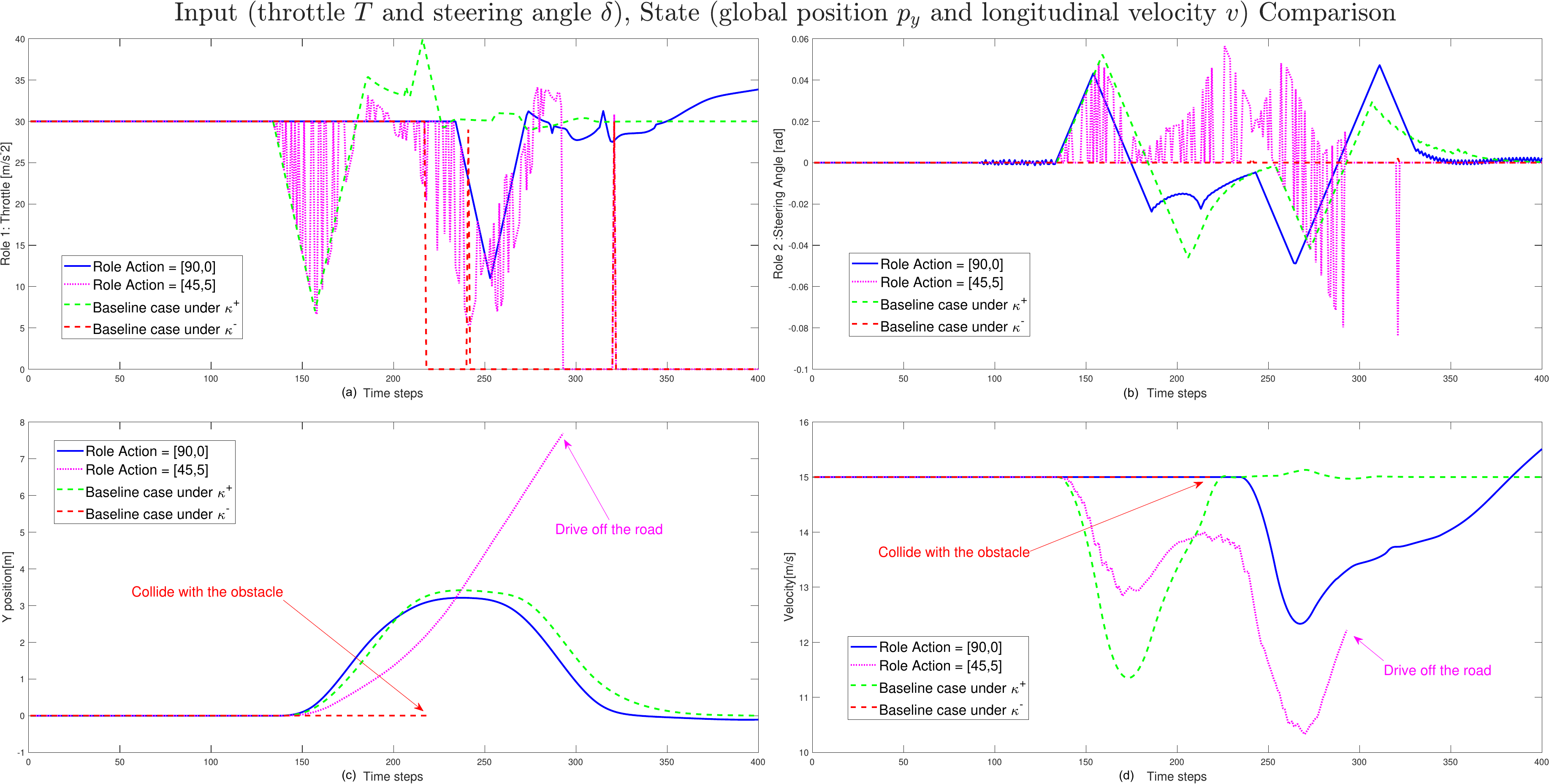}
    \caption{Input $T,\delta$ and states $p_y, v$ comparison between SCG Case 1 and baseline cases. (a): throttle $T$. (b): steering angle $\delta$. (c): position in y axis $p_y$. (d): velocity $v$.}
    \label{fig5}
\end{figure}
\begin{figure}[thpb]
    \centering
    \includegraphics[width=0.48\textwidth]{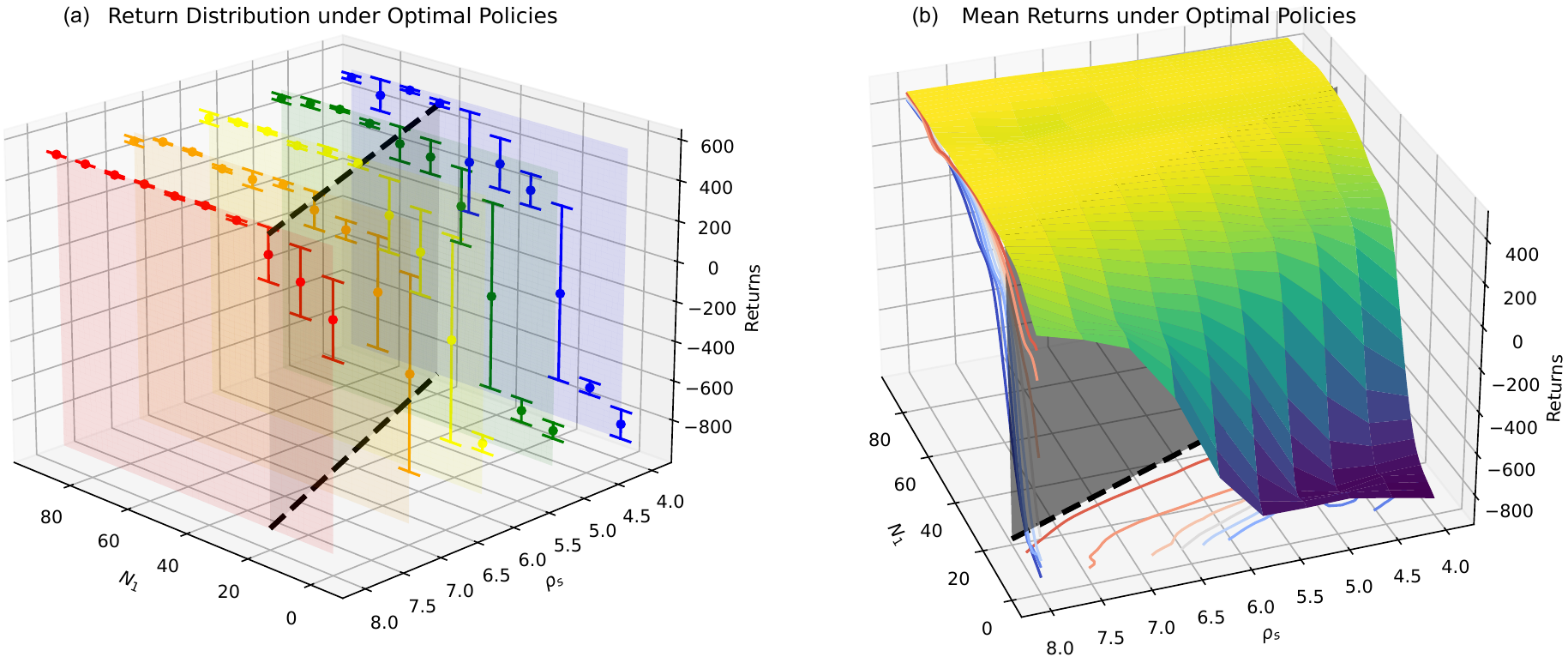}
    \caption{With $N_2$ is set constant $N_2=10$, we study cases with different settings: number of players in group 1, $N_1\in[0,90]$ and social role of players in group 2, $\rho_s\in[4,8]$. (a): Return distributions under optimal policies for each case, where mean, max and min return are marked. (b): mean returns for each case. The black dashed lines define the boundary between cases with and without CI.}
    \label{fig6}
\end{figure}

Now we specify the configurations for the numerical simulation as Table \ref{table1} outlines. We simulate scenarios with $N$ players, divided into two groups: $N_1$ with $\tau t_s$ delayed information and $N_2$ without delays. Players collectively select roles at intervals of $K$. The MPC controller is sampled by $t_s$ where prediction and control horizons is denoted as $N_p$ and $N_c$, respectively, with weight matrices $Q$ and $R$. Additional parameters include car dimensions $W,L,l_f,l_r$, acceleration coefficients $\alpha{}$, learning rates for actor and critic networks $a\_lr, c\_lr$, and the discount factor $\gamma$. The degree of the $\mathcal{G}_k$ nodes is limited between $d_{min}$ and $d_{max}$. Specifically, we mainly consider three cases: Case 1: $N_1, N_2, \rho_s = 90, 10, 8$. Case 2: $N_1, N_2, \rho_s = 90, 10, 4$. Case 3: $N_1, N_2, \rho_s = 10, 10, 8$.

The car is driving on the center lane of a road with three 6-meter-width lanes at a constant velocity $15m/s$ without any inputs from time step $k = 0$ to $k=100(t=2s)$, states evolving from $[0, 0, 0, 15]^T$ to
$[30, 0, 0, 15]^T$, which sets initial conditions for the avatar car. An obstacle, measuring 5 meters in length and 2 meters in width, is positioned at $(p_x = 100, p_y = 0)$ and becomes observable when the car approaches within 30 meters of its position.

The reward function $R = R_1 + R_2$. During each interval of $Kt_s$, $R_1 = -(15 - v_{\text{avg}})^2$, where $v_{\text{avg}}$ represents the average velocity of the car. If $v_{\text{avg}} \geq 15$, then $R_1 = 0$. Additionally, $R_2$ is set to $-500$ if the car collides with the obstacle, or goes off the road, or if it fails to pass the obstacle within 400 steps (i.e., 8 seconds) without violating any constraints. Otherwise, $R_2 = 500$.


\subsection{Simulation Results}

As depicted in Fig. 3 (a), for Case 1, the social roles of players in these two groups converge separately, promoting returns as the learning process progresses, eventually establishing a DOL where the group with information delays solely controls the throttle while the other solely controls the steering. The expected returns ultimately converge to $J_0$. This demonstrates the emergence of CI through the learning of social roles. As shown in Fig. 4, players initially fail to avoid collisions, as labeled as $\underline{a}=[45,5]$, yet eventually transitioning to $\underline{a}=[90,0]$, performing as effectively as the Baseline case under $\kappa^+$. We can observe the decision and state evolution for both $\underline{a}=[45,5]$ and $\underline{a}=[90,0]$ in Fig. 5. 

We conducted comparison simulations for Case 2 and Case 3. As depicted in Fig. 3 (a), the establishment of division of labor occurs much slower in these cases. Moreover, theoretical analysis reveals that even after the division of labor is established, neither case is capable of reaching the performance level $J_0$ in Fig. 3 (b). To provide further insight, we examined additional cases with varying values of $N_1$ and $\rho_s$, showcasing the return distributions under optimal policies for each case in Fig. \ref{fig6} (a) and the mean returns in Fig. \ref{fig6} (b). As the black dashed line, i.e., $\rho_s=-0.1N_1+10$ in Fig. \ref{fig6} (a) shows, the emergence of CI requires: $\rho_s\geq-0.1N_1+10$, which further implies : (a) $\frac{N_1 + \rho_s N_2}{K}\geq1.25$, and (b) $\frac{\rho_sN_2}{N_1}\geq11.1\%$. 

Given that the external stimulus remains constant across these cases, our findings justify the theoretical results in \textit{Remark 8} suggests. Furthermore, we notice it is always taken for granted that society comprised solely of elite individuals would be the most efficient and productive, largely due to the emphasis on condition (b). However, interestingly, our work underscores the significance of commoners, urging equal attention to condition (a).  

\section{CONCLUSIONS}
\label{sec5}

By investigating LinYi's Experiments, this paper offered crucial insights into the factors fostering CI. Through both simulations and theoretical analysis, we found the emergence of CI relies on the learning of appropriate social roles, and specific inherent factors regarding player numbers, their distribution, and the allocation of social power. Remarkably, we also disclosed a counter-intuitive finding: CI cannot emerge in a pure-elite society without inclusion of commoners. In our future work, we will conduct human experiments on this SCG to further validate our findings.


\addtolength{\textheight}{-12cm}   





\end{document}